\documentclass[fleqn,twoside]{article}
\usepackage{espcrc2}
\usepackage{pifont}
\usepackage{graphicx}
\usepackage[figuresright]{rotating}

\newcommand{\AmS}{{\protect\the\textfont2
  A\kern-.1667em\lower.5ex\hbox{M}\kern-.125emS}}

\title{Double real radiation corrections to gluon scattering at NNLO}

\author{Joao Pires\address{Institute for Particle Physics Phenomenology\\
	Department of Physics\\
	University of Durham\\
	DH1 3LE\\ 
	UK}%
        \thanks{This research was supported in part by the UK Science and Technology Facilities
Council and by the European Commission's Marie-Curie Research Training Network
under contract MRTN-CT-2006-035505 `Tools and Precision Calculations for Physics
Discoveries at Colliders'. EWNG gratefully acknowledges the support of the
Wolfson Foundation and the Royal Society. JP gratefully acknowledges the award
of a Funda\c{c}\~{a}o para a Ci\^encia e Tecnologia (FCT - Portugal) PhD
studentship.},
        E.W.N. Glover\addressmark,
        }
       
\begin{document}

\begin{abstract}
We use the antenna subtraction method to isolate the double real radiation infrared singularities present in the six-gluon tree-level process at next-to-next-to-leading order.  We show numerically that the subtraction term correctly approximates the matrix elements in the various single and double unresolved configurations.
\vspace{1pc}
\end{abstract}

\maketitle

\section{Introduction}
In proton-proton collisions, the factorised form of the inclusive cross section is given by,
\begin{equation}
{\rm d}\sigma =\sum_{i,j} \int   
{\rm d}\hat\sigma_{ij}
 f_i(\xi_1) f_j(\xi_2) \frac{d\xi_1}{\xi_1} \frac{d\xi_2}{\xi_2} \nonumber
\end{equation}
where ${\rm d}\hat\sigma_{ij}$ is the parton-level scattering cross section for parton $i$ to scatter off parton $j$ normalised to the hadron-hadron flux\footnote{The partonic cross section normalised to the parton-parton flux is obtained by absorbing the inverse factors of $\xi_1$ and $\xi_2$ into ${\rm d}\hat\sigma_{ij}$.} and the sum runs over the possible parton types $i$ and $j$. The probability of finding a parton of type $i$ in the proton carrying a momentum fraction $\xi$ is described by the parton distribution function $f_i(\xi)d\xi$.  By applying suitable cuts, one can study more exclusive observables such as the transverse momentum distribution or rapidity distributions of the hard objects (jets or vector bosons, higgs bosons or other new particles) produced in the hard scattering.  The leading-order (LO) prediction is a useful guide to the rough size of the cross section, but is usually subject to large uncertainties from the dependence on the unphysical renormalisation and factorisation scales, as well as possible mismatches between the (theoretical) parton-level and the (experimental) hadron-level.  

The theoretical prediction may be improved by including 
higher order perturbative predictions which have the effect of (a) reducing the renormalisation/factorisation scale dependence and (b) improving the matching of the parton level event topology with the experimentally observed hadronic final state~\cite{Glover:2002gz}.
The partonic cross section ${\rm d}\hat\sigma_{ij}$ has the perturbative expansion 
\begin{eqnarray}
\lefteqn{{\rm d}\hat\sigma_{ij}={\rm d}\hat\sigma_{ij}^{LO}
+\left(\frac{\alpha_s}{2\pi}\right){\rm d}\hat\sigma_{ij}^{NLO}
+\left(\frac{\alpha_s}{2\pi}\right)^2{\rm d}\hat\sigma_{ij}^{NNLO}}\nonumber \\
&&
+{\cal O}(\alpha_s^3)
\end{eqnarray}
where the next-to-leading order (NLO) and next-to-next-to-leading order (NNLO) strong corrections are identified. For many processes, knowledge of the NLO correction is sufficient.   However, for the main $2 \to1$ or $2 \to 2$ scattering processes such as Drell-Yan, Higgs production, di-jet production,
vector-boson plus jet, vector-boson pair production or heavy quark pair production,  the NLO corrections   
still have a large theoretical uncertainty and it is necessary to include the
NNLO perturbative corrections.  In addition to reducing the renormalisation and factorisation scale dependence 
there is an improved matching of the parton level theoretical jet algorithm with the hadron level
experimental jet algorithm because the jet structure can be modeled by the
presence of a third parton.  In this talk, we are mainly concerned with the NNLO corrections to di-jet production 
where the resulting theoretical uncertainty
is estimated to be at the few per-cent level~\cite{Glover:2002gz}.

\begin{center}
\begin{table*}[tbh]
\caption{Integrated antenna functions}
\label{table:1}
\renewcommand{\tabcolsep}{2pc} % enlarge column spacing
\renewcommand{\arraystretch}{1.2} % enlarge line spacing
\begin{tabular}{|c|c|c|c|}
\hline
         & {\rm final-final} & {\rm final-initial} & {\rm initial-initial}   \\
\hline
$X_3^0$                &  \ding{51}\cite{GehrmannDeRidder:2005cm} & \ding{51}\cite{Daleo:2006xa}  &  \ding{51}\cite{Daleo:2006xa}     \\
$X_4^0$                &  \ding{51}\cite{GehrmannDeRidder:2005cm} & \ding{51}\cite{Daleo:2009yj}  &   (in progress) \cite{Boughezal:2010ty}  \\
$X_3^0\otimes X_3^0$   &  \ding{51}\cite{GehrmannDeRidder:2005cm} &  (in progress) &   (in progress)  \\
$X_3^1$  	       & \ding{51}\cite{GehrmannDeRidder:2005cm} &  \ding{51}\cite{Daleo:2009yj} &  (in progress)    \\
\hline
\end{tabular}\\[2pt]
\end{table*}
\end{center}

\section{Antenna subtraction}

Any calculation of these higher-order corrections requires a systematic procedure for extracting the infrared singularities that arise when one or more final state particles become soft and/or collinear.  These singularities are present in the real radiation contribution at next-to-leading order (NLO), and in double real radiation and mixed real-virtual contributions at next-to-next-to-leading order (NNLO). 

There have been several approaches to build a general subtraction scheme at NNLO \cite{GehrmannDeRidder:2005cm,Weinzierl:2003fx,Frixione:2004is,Somogyi:2005xz,Somogyi:2006da,Somogyi:2006db,Somogyi:2008fc,Aglietti:2008fe,Somogyi:2009ri,Bolzoni:2009ye,Czakon}. 
We will follow the NNLO antenna subtraction method which was derived in \cite{GehrmannDeRidder:2005cm} for processes involving only (massless) final state partons. This formalism has been applied in the computation of NNLO corrections to three-jet production in electron-positron annihilation \cite{GehrmannDeRidder:2007jk,GehrmannDeRidder:2008ug,Weinzierl:2008iv,Weinzierl:2009nz} and related event shapes \cite{GehrmannDeRidder:2007bj,GehrmannDeRidder:2007hr,GehrmannDeRidder:2009dp,Weinzierl:2009ms,Weinzierl:2009yz}.  It has also been extended at NNLO to include one hadron in the initial state relevant for electron-proton scattering~\cite{Daleo:2009yj,Daleo:2010tz} while in refs.~\cite{Boughezal:2010ty,GloverPires} the extension of the antenna subtraction method to include two hadrons in the initial state is discussed. 

In the antenna subtraction method the subtraction term is constructed from products of 
antenna functions with reduced matrix elements (with fewer final state 
partons than the original matrix element), and integrated over 
a phase space which is factorised into an antenna phase space (involving all
unresolved partons and the two radiators) multiplied with a reduced phase 
space (where the momenta of radiators and unresolved radiation are replaced 
by two redefined momenta).
The full subtraction term is obtained by summing over all antennae 
required for the problem under consideration. In the most general case
(two partons in the initial state, and two or more hard partons in the final 
state), this sum includes final-final, initial-final and initial-initial 
antennae.  

The relevant antenna is determined by both the external state and the pair of hard
partons it collapses to.   In general we denote the antenna function as $X$.
For antennae that collapse onto a hard quark-antiquark pair, 
$X = A$ for $qg \bar q$.  Similarly, for quark-gluon antenna, we have 
$X = D$ for $qg g$ and $X=E$ for $qq^\prime
\bar q^\prime$ final states.  Finally, we characterise the gluon-gluon antennae as
$X=F$ for $ggg$,  $X=G$ for $gq\bar q$ final states.
At NNLO we will need four-particle antennae involving two unresolved partons and one-loop three-particle antennae.

In all cases the antenna functions are derived from physical matrix elements: 
the quark-antiquark antenna functions from 
$\gamma^* \to q\bar q~+$~(partons)~\cite{GehrmannDeRidder:2004tv}, the quark-gluon antenna 
functions from $\tilde\chi \to \tilde g~+$~(partons)~\cite{GehrmannDeRidder:2005hi} and 
the gluon-gluon antenna functions from $H\to$~(partons)~\cite{GehrmannDeRidder:2005aw}.  

A key element in any subtraction method, is the ability to add back the subtraction term integrated over the unresolved phase space. In the antenna approach, this integration needs to be performed once for each antenna.  Although no problem in principle has been identified, at present, not all of the necessary integrals have been completed.  The current state of play is summarized in Table~1.

\section{Double real radiation gluonic contributions}

We now consider the six-gluon contribution to the NNLO di-jet cross section.   At leading colour\footnote{To simplify the discussion, we systematically drop the sub-leading colour terms.}, the double real radiation contribution is given by,
\begin{eqnarray}
\lefteqn{{\rm d}\hat\sigma_{NNLO}^R =  {\cal N} \left(\frac{\alpha_s N}{2\pi}\right)^2
\sum_{\sigma\in S_6/Z_6} A_{6}^0 (\sigma(1),\ldots,\sigma(6))}\nonumber \\
&\times &\frac{1}{4!}{\rm d}\Phi_{4}(p_{3},\ldots,p_{6};p_1,p_2)\, J_{2}^{(4)}(p_{3},\ldots,p_{6}).\label{eq:6gNNLO} 
\end{eqnarray}
where the normalisation factor ${\cal N}$ includes all QCD-independent factors as well as the LO dependence on the renormalised QCD coupling constant $\alpha_s$. $S_n/Z_n$ is the group of non-cyclic permutations of $n$ symbols and $A_{6}^0 (\sigma(1),\ldots,\sigma(6))$ is the square of a colour ordered amplitude. 
$p_1$ and $p_2$ are the momenta of the ingoing gluons, while $p_3,\ldots,p_6$ are the momenta of the four final state gluons and where the phase space integration ${\rm d}\Phi_{4}$ is over the $4$-parton final state provided that precisely $2$-jets are observed by the jet algorithm $J_2^{(4)}$.
Here singularities occur in the phase space regions corresponding to two gluons becoming simultaneously soft and/or collinear and the subtraction term must successfully regularize all double unresolved and single unresolved singularities.

The antenna subtraction term ${\rm d}\hat\sigma_{NNLO}^S$ for the six-gluon process was derived in Ref.~\cite{GloverPires}.

\section{Numerical checks}

In this section we will test how well the subtraction term ${\rm d}\hat\sigma^S_{NNLO}$ approaches the 
double real contribution ${\rm d}\hat\sigma^R_{NNLO}$ in the single and double unresolved regions of phase space so that their difference can be integrated numerically over the four-dimensional phase space.
We will do this numerically by generating a series of phase space points that approach a given double or single unresolved limit. For each generated point we compute the ratio,
\begin{equation}
R=\frac{{\rm d}\hat\sigma^R_{NNLO}}{{\rm d}\hat\sigma^S_{NNLO}}
\end{equation}
where ${\rm d}\hat\sigma^R_{NNLO}$ is the matrix element squared given in equation (\ref{eq:6gNNLO}) and 
${\rm d}\hat\sigma^S_{NNLO}$ is the subtraction term of Ref.~\cite{GloverPires}.  
The ratio of the matrix element and the subtraction term should approach unity as we get closer to any singularity and the quality of this convergence represents the first sanity check on the NNLO antenna
subtraction method for the six-gluon  process.  

For each unresolved configuration, we will define a variable that controls how we approach the singularity subject to the requirement that there are at least two jets in the final state with $p_T>$50 GeV. The centre-of-mass energy $\sqrt s$ is fixed to be 1000~GeV.
All of the single and double unresolved
regions of   phase space were analysed separately. In each case,  we generated 10000 random phase space points and computed the average value of $R$ together with the standard deviation. 
The results of the numerical study for all limits (apart from the single collinear limit we shall discuss below) are summarized in Table~2.

The numerical results show that the combination of the antennae in the subtraction term correctly describes the infrared singularity
structure of the matrix element. All of the double
unresolved and single soft singularities present in the matrix elements are cancelled
in a point-by point manner.  

\begin{center}
\begin{table*}[tbh]
\caption{Results for the average and standard deviation of $R$ for 10000 phase space points close to singular limits.  }
\label{table:2}
\renewcommand{\arraystretch}{1.2} % enlarge line spacing
\begin{tabular}{|l|l|l|l|}
\hline
         & {\rm small parameter}  & $<R>$  & $\sigma$ \\
\hline
double soft                & $ (s -s_{ij})/s =10^{-6}$  &     $0.9999994$ & $4\times10^{-5}$  \\
triple collinear (final)   & $ s_{ijk}/s =10^{-9}$  &    $1.0000004$ & $4.2\times10^{-5}$  \\
triple collinear (initial) & $ s_{1jk}/s =-10^{-10}$  &  $0.99954$   & $0.04$  \\
soft/collinear (final)     & $ (s -s_{ijk})/s =10^{-6}$, $ s_{ij}/s =10^{-6}$  &   $0.99999993$  & $0.0001$  \\
soft/collinear (initial)   & $ (s -s_{ijk})/s =10^{-7}$, $ s_{1i}/s =-10^{-7}$     & $0.99999998$  &  $1.6\times10^{-7}$ \\
double collinear (final-final)   &  $ s_{ij}/s =10^{-8}$, $ s_{kl}/s =10^{-8}$ &   $0.9999995$  &  $0.00037$ \\
double collinear (final-initial)   &  $ s_{jk}/s =10^{-10}$, $ s_{1i}/s =-10^{-10}$ &   $1.00012$  &  $0.018$ \\
double collinear (initial-initial) &  $ s_{1i}/s =-10^{-10}$, $ s_{2j}/s =-10^{-10}$  & $1.00001$  &  $0.004$ \\
single soft                &  $ (s -s_{ijk})/s =10^{-7}$ &   $0.999998$ &  $1.9\times10^{-5}$ \\
\hline
\end{tabular}\\[2pt]
\end{table*}
\end{center}

\begin{figure}[th!]
\begin{center}
\includegraphics[width=4.2cm,angle=270]{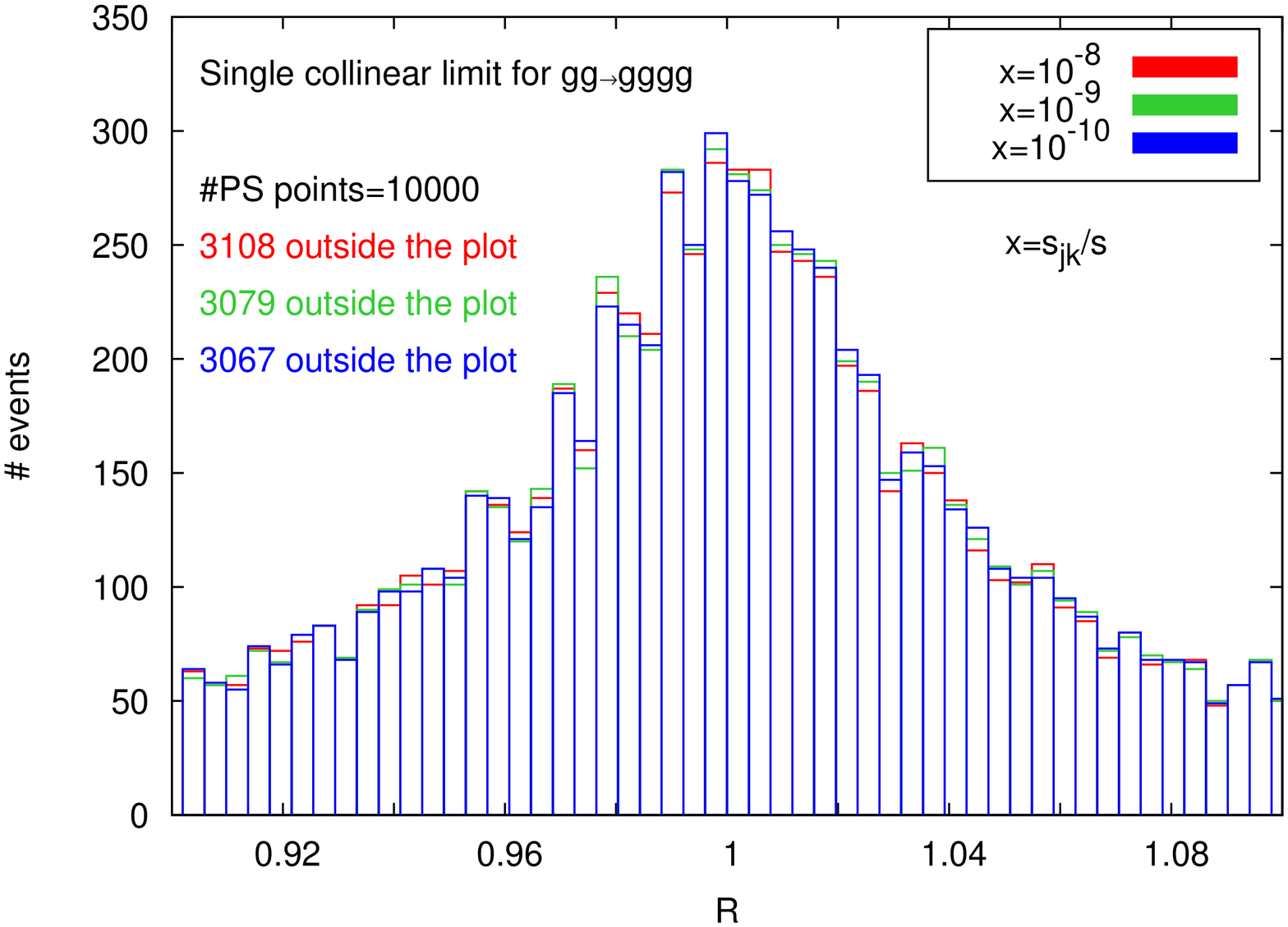}\\
 \includegraphics[width=4.2cm,angle=270]{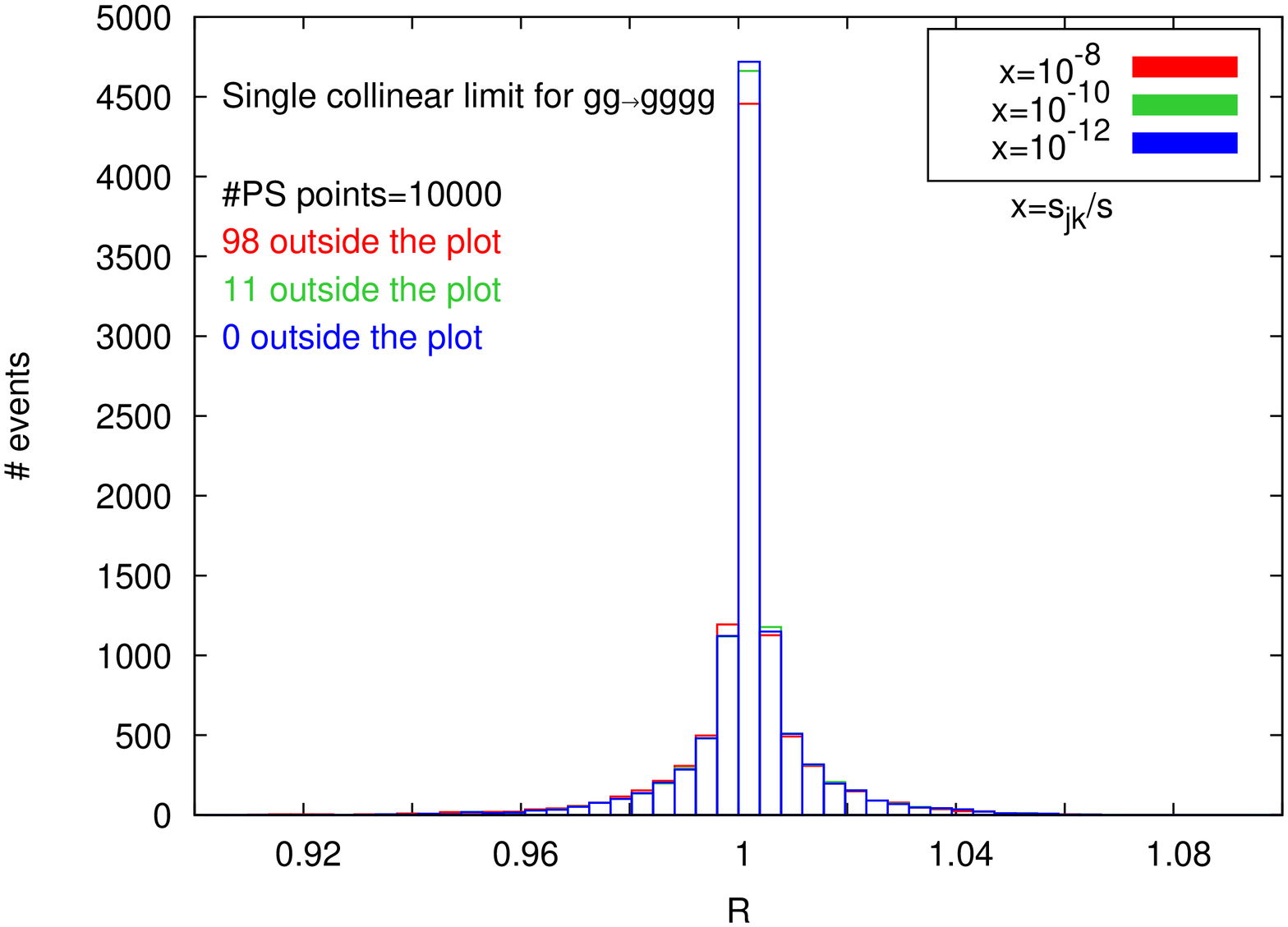}\\
 \includegraphics[width=4.2cm,angle=270]{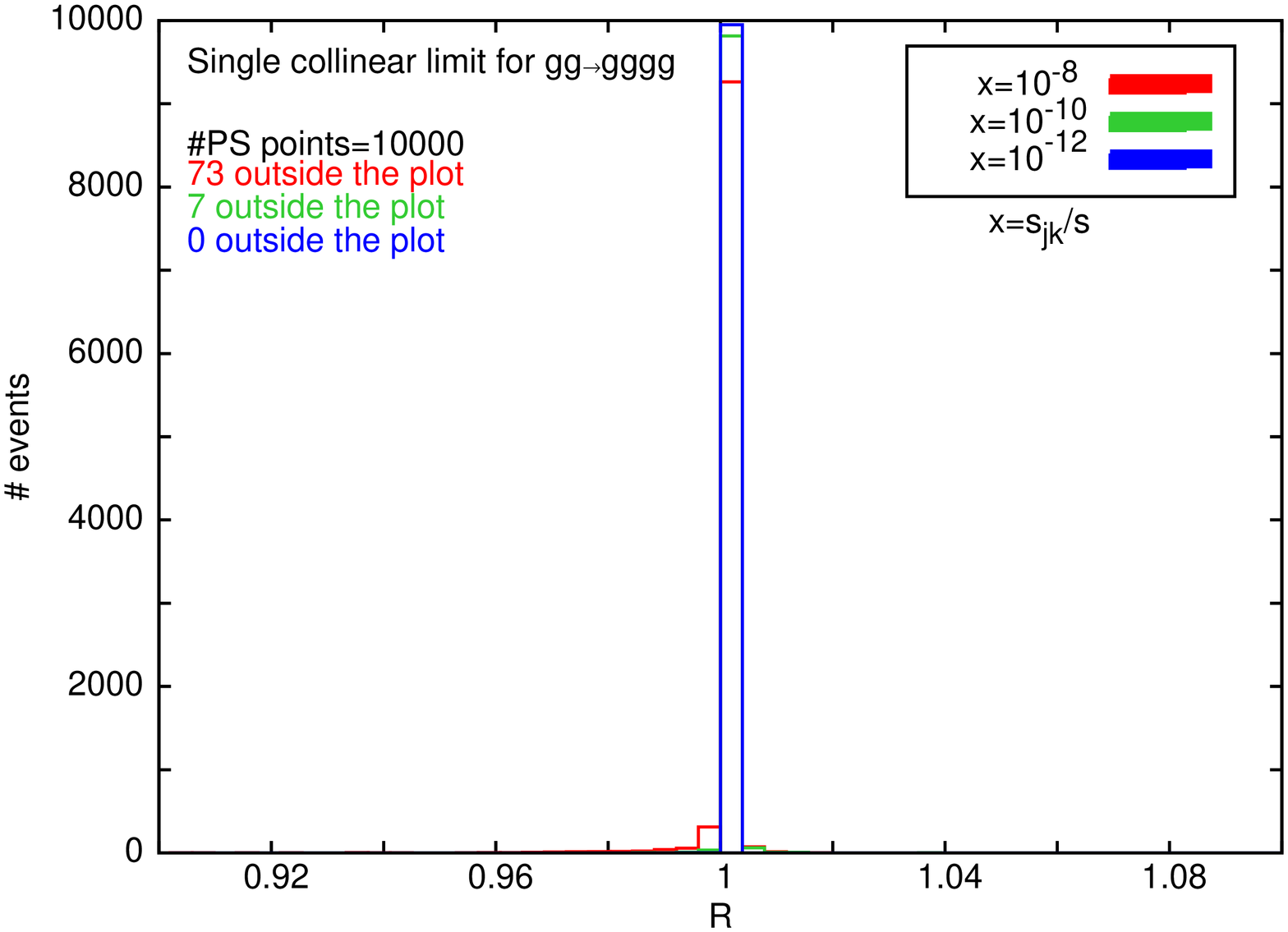}\\
 \end{center}
\caption[Azimuthally corrected single collinear limit]{Distribution of $R$ for 10000 single final state collinear phase space points for 
$x = s_{ij}/s $ and $x=10^{-8}$ (red),   $x=10^{-9}$ (green) and  $x=10^{-10}$ (blue) for (a) the uncorrected matrix subtraction term (b) the azimuthally corrected subtraction term and (c) pairs of phase space points related by a rotation of $\pi/2$ about the axis of the collinear pair.}
\label{fig:scollff}
\end{figure}

\begin{figure}[th!]
\begin{center}
\includegraphics[width=4.2cm,angle=270]{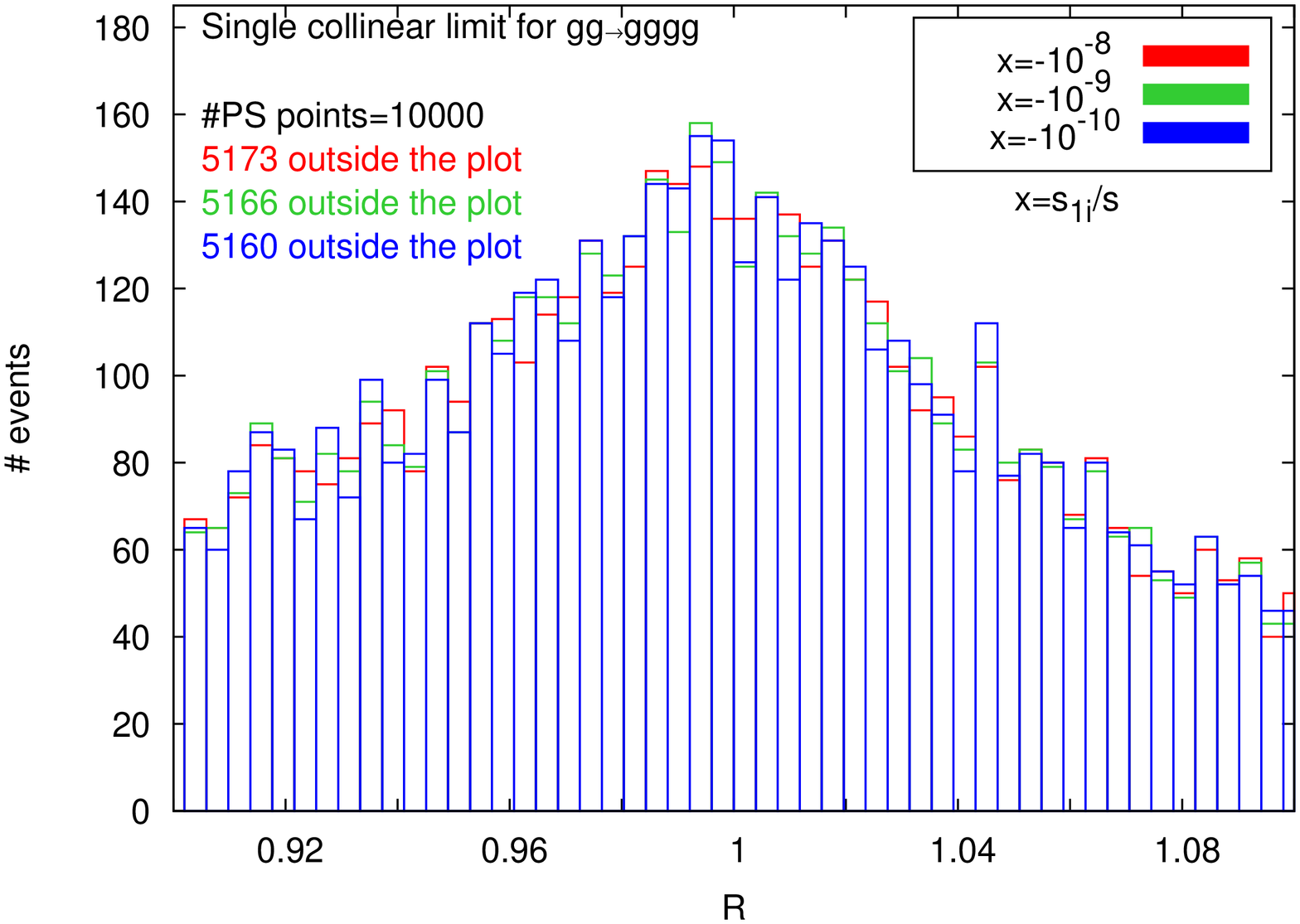}\\
 \includegraphics[width=4.2cm,angle=270]{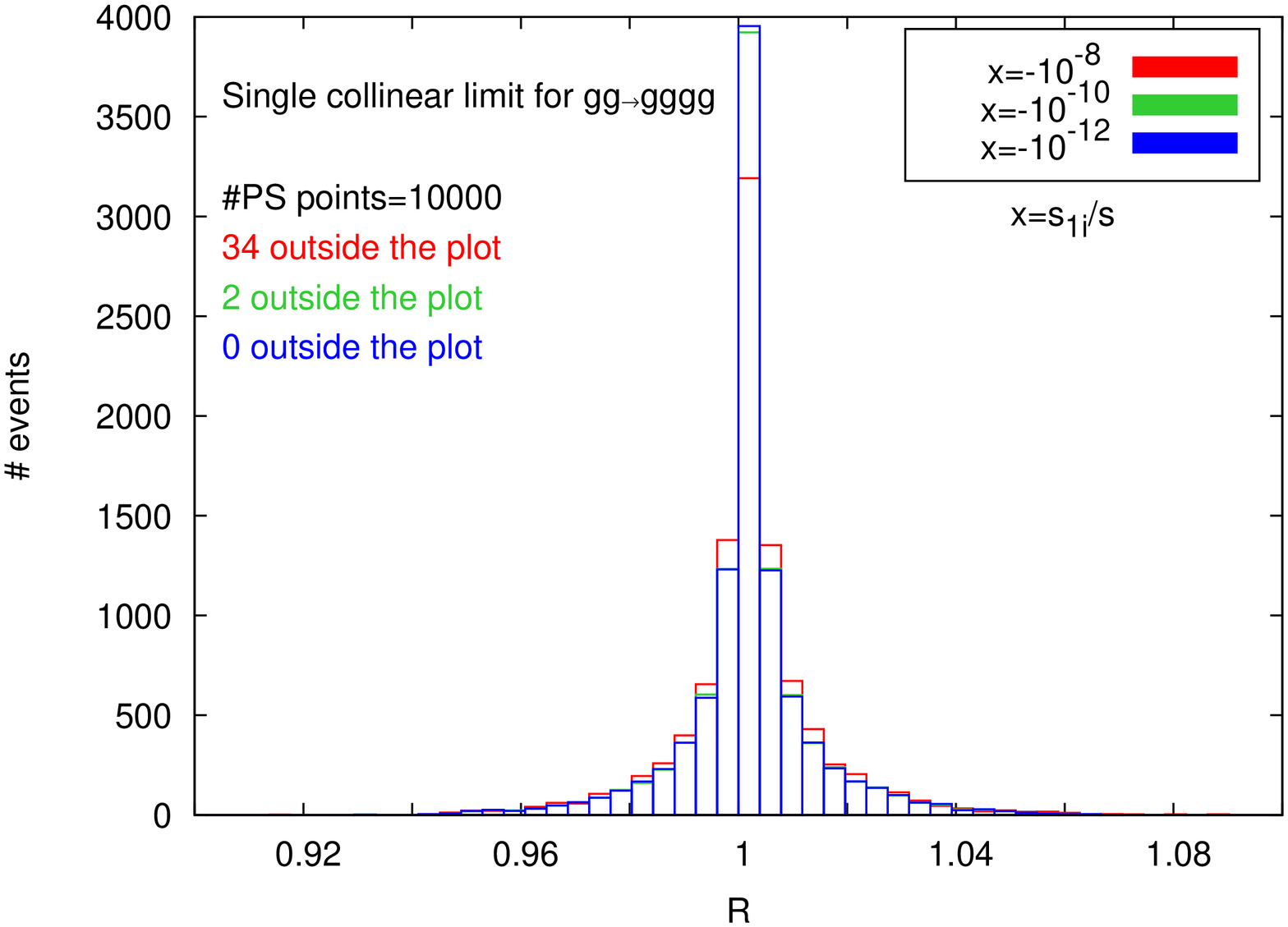}\\
 \includegraphics[width=4.2cm,angle=270]{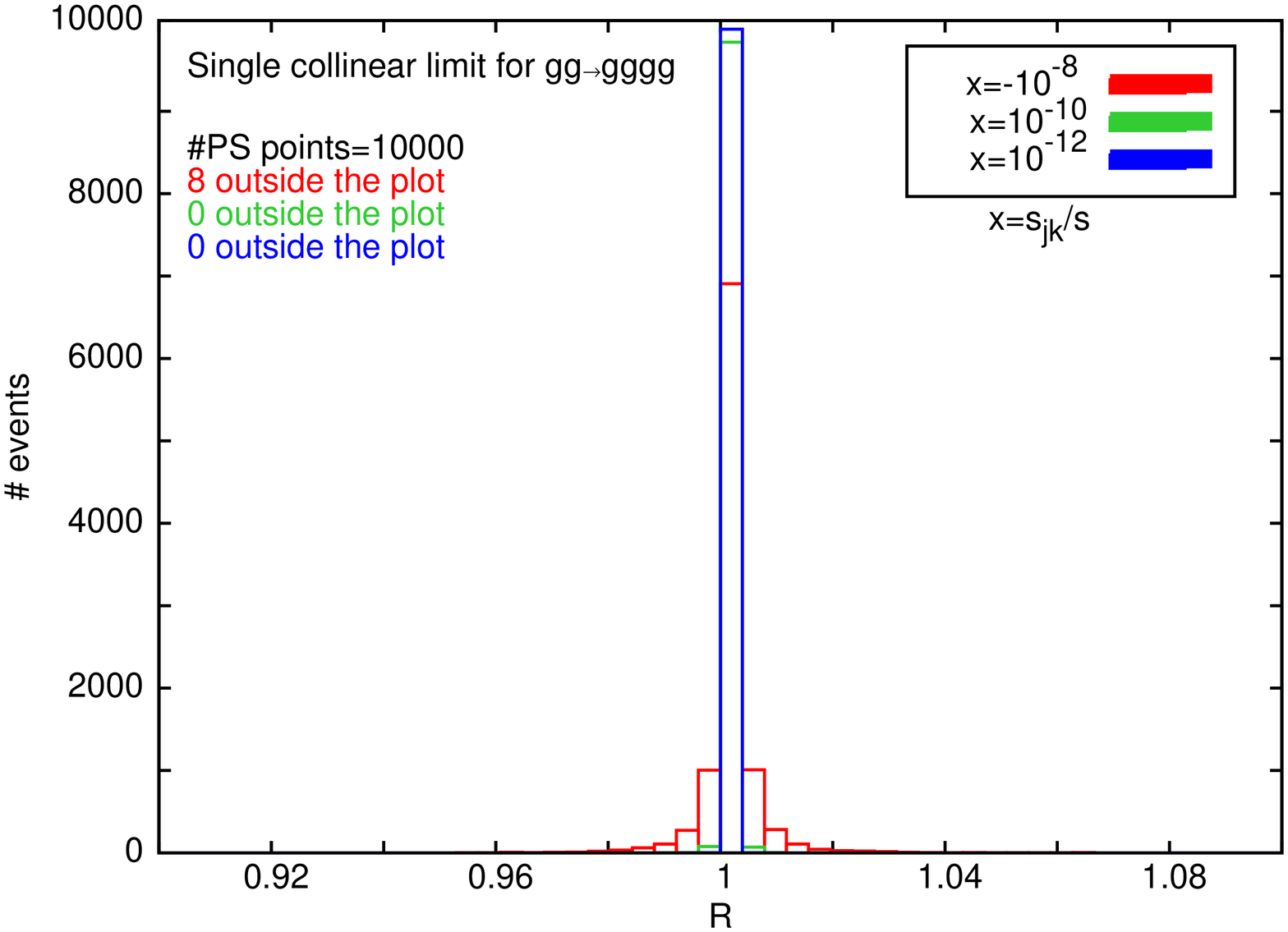}\\
 \end{center}
\caption[Azimuthally corrected single collinear limit]{Distribution of $R$ for 10000 single initial state collinear phase space points for (a) the uncorrected matrix subtraction term (b) the azimuthally corrected subtraction term and (c) pairs of phase space points related by a rotation of $\pi/2$ about the axis of the collinear pair and then boosted so the initial direction is preserved.}
\label{fig:scollif}
\end{figure}

Finally we generate points corresponding to the final and initial state single collinear regions of the phase space.  
For the final-final collinear singularity,  fig.~\ref{fig:scollff}(a) shows the $R$-distribution for a range of values of $x = s_{ij}/s $. 
Similarly in the initial-final collinear limit, we show the distributions of $R$ for small values of $x = s_{1i}/s $ in fig.~\ref{fig:scollif}(a).
We see that the subtraction term, which is based on azimuthally averaged antennae functions does not accurately describe the azimuthal correlations present in the matrix elements and antenna functions when a gluon splits into two collinear gluons.   This is the reason why the distributions in figs.~\ref{fig:scollff}(a) and \ref{fig:scollif}(a) have such a broad shape.  It is clear that as we approach the collinear limits $x \to 0$,
the azimuthal terms are not suppressed and the subtraction term is not, point by point, an adequate representation of the matrix element.

Nevertheless, the azimuthal terms coming from the single collinear limits do vanish after an azimuthal integration over the unresolved phase space. Here we are performing a point-by-point analysis on the integrand defined by the matrix element squared and the subtraction term. One possible strategy is to introduce a correction term $\Theta_{F_3^0}(i,j,z,k_\perp)$ to the $F_4^0$ four-gluon antenna functions which reconstructs the angular terms~\cite{GloverPires}.
 Subtracting $\Theta_{F_3^0}(i,j,z,k_\perp)$  from the  final-final  and, initial-final and initial-initial configurations (by crossing momenta to the initial state) produces a subtraction term that is locally free of angular terms.

With this azimuthally modified subtraction term, we recompute the distributions in figs.~\ref{fig:scollff}(b) and ~\ref{fig:scollif}(b).  For 10000
final state single collinear phase space points and $x=10^{-12}$ we obtained an average $R=0.99994$ and a standard deviation of $\sigma=0.015$. We
repeated the same analysis for the initial state collinear configuration for $x=-10^{-12}$ and obtained an average of $R=1.00007$ and a standard
deviation of $\sigma=0.012$. For both cases, the distributions now peak around $R=1$ with a more pronounced peak as the limit is approached, just as
in the double unresolved and single soft limits discussed earlier.  This demonstrates the convergence of the counterterm to the matrix element. 

However, the azimuthal correction term $\Theta_{F_3^0}(i,j,z,k_\perp)$ has the unfortunate side effect of generating new divergences which are not present in the matrix element. 

A second more successful approach is to cancel the angular terms by combining phase space points related to each other by a rotation of the system of unresolved partons \cite{Weinzierl:2006wi,GehrmannDeRidder:2007jk}.  When both collinear partons are in the final state, this can be achieved by considering pairs of phase space points which are related by rotating the collinear partons by an angle of $\pi/2$ around the resultant parton direction. 
Similarly, for the initial-final state collinear configurations produced when $p_i^\mu \to p^\mu+p_j^\mu$ for $i=1,2$ and with $i||j$ and $p^2=0$,  the phase space points should again be related by rotations of $\pi/2$ about the direction of $p^\mu$.  This has the consequence of rotating $p_i^\mu$ off the beam axis and therefore has to be compensated by a Lorentz boost.

The effect of combining pairs of phase space points is shown in Figs.~\ref{fig:scollff}(c) and ~\ref{fig:scollif}(c). We see that the distributions for both final-final and initial-final collinear limits have a very sharp peak around $R=1$.   For the final state singularity and $x=10^{-12}$ we obtained an average of $R=0.999996$ and a standard deviation of $\sigma=0.00015$. Similarly, for the 
initial state collinear singularity and $x=-10^{-12}$ we found $R=0.9999991$ and $\sigma=0.00011$.
There is an enormous improvement compared to the raw distributions shown in Figs.~\ref{fig:scollff}(a) and \ref{fig:scollif}(a) and a significant improvement compared to adding an azimuthal correction term shown in Figs.~\ref{fig:scollff}(b) and \ref{fig:scollif}(b).  The phase space point averaged subtraction term clearly converges to the matrix element as we approach the single collinear limit and correctly subtracts the azimuthally enhanced terms in a point-by-point manner.

\section{Conclusions}

In this contribution, we have discussed the application of the antenna
subtraction formalism to construct the subtraction term relevant for the six-gluon
double real radiation contribution to di-jet production. The subtraction term is
constructed using four-parton and three-parton antennae.  We showed that the
subtraction term correctly describes the double unresolved limits of the $gg \to
gggg$ process.    In particular, by combining phase space points, the subtraction term avoids the problems associated with angular correlations produced when a gluon splits into two collinear gluons.  
The final goal is the construction of a numerical program to compute
NNLO QCD estimates of di-jet production in hadron-hadron collisions.

\end{document}